\newcommand{\gray}{$\gamma$-ray\xspace}
\newcommand{\nray}{$\nu$\xspace}
\newcommand{\grays}{$\gamma$-rays\xspace}
\documentclass{webofc}
\usepackage[varg]{txfonts}   
\woctitle{The Roma International Conference on Astroparticle Physics}
\begin{document}
\title{Galactic sources of high energy neutrinos:}
%
\subtitle{Expectation from gamma-ray data}

\author{N. Sahakyan \inst{1}\fnsep\thanks{\email{narek@icra.it}} 
}

\institute{ICRANet-Yerevan, Marshall Baghramian Avenue 24a, Yerevan 0019, Republic of Armenia
          }

\abstract{%
 The recent results from ground based \gray detectors (HESS, MAGIC, VERITAS) provide a population of TeV galactic \gray sources which are potential sources of High Energy (HE) neutrinos. Since the \grays and \nray -s are produced from decays of neutral and charged pions, the flux of TeV \grays can be used to estimate the upper limit of \nray flux and vice versa; the detectability of \nray flux implies a minimum flux of the accompanying \grays (assuming the internal and the external absorption of \grays is negligible). Using this minimum flux, it is possible to find the sources which can be detected with cubic-kilometer telescopes. I will discuss the possibility to detect HE neutrinos from powerful galactic accelerators, such as Supernova Remnants (SNRs) and Pulsar Wind Nebulae (PWNe) and show that likely only RX J1713.7-3946 , RX J0852.0-4622 and Vela X can be detected by current generation of instruments (IceCube and Km3Net). It will be shown also, that galactic binary systems could be promising sources of HE \nray -s. In particular, \nray -s and \grays from Cygnus X-3 will be discussed during recent gamma-ray activity, showing that in the future such kind of activities could produce detectable flux of HE \nray -s.
}
\maketitle
\section{Introduction}
\label{intro}
The neutrinos are unique messengers which carry astrophysical/cosmological information \cite{ahar1}. Since they can escape from much denser environments (high photon field and/or matter) than \grays, it provides a possibility to examine the sources which are impossible by \gray detectors. The detection of HE \nray -s will help to understand the origin of cosmic rays since they are produced in hadronic interactions, thus bring direct information on the hadronic component accelerated in the sources. Many galactic and extragalactcic astrophysical object classes are able to accelerate particles above TeV energies which are confirmed by the recent observations with ground based \gray detectors (HEGRA, HESS, MAGIC, VERITAS and etc). While the acceleration of leptons (electrons) is limited by relatively short cooling times, the acceleration of hadrons goes beyond TeV/PeV energies. The interaction of these hadrons (with photon fields or matter) produces HE neutrinos ($>$ TeV). Recently, the IceCube detector has detected PeV neutrinos \cite{icpev}, which are probably produced in extragalactic objects (e.g. active galactic nuclei, the source of gamma-ray burst and etc). However, the  galactic sources (especially supernova remnants) have been historically considered more favorable  sources for HE neutrino detectors. This is justified considering that the galactic sources are responsible for acceleration of cosmic rays at energies below the so-called knee. Perhaps, the chance to detect the first galactic sources will increase with the arrival of KM3Net detector considering the northern hemisphere is more suitable to study galactic sources \cite{viss}.\\
Current knowledge on the population and intensity of galactic nonthermal sources (provided by \gray detectors) allows to make quantitative estimation on the expectations for HE \nray detectors. Indeed, \grays and neutrinos are produced almost in the same rate in $pp$ interaction, thus \gray data can be used as a guide for neutrino searches. In \cite{viss}, it was shown that the most relevant energy range for \gray observations (which will provide one event in neutrino detector) is around 20 TeV, where the intensity of \gray is at the narrow range of $I_\gamma\sim(2-6)\times10^{-15}\:TeV^{-1}\:cm^{-2}\:s^{-1}$ \cite{viss}. However, for now only limited data are available at 20 TeV, which will be highly explored with the future planned CTA instrument, therefore we try to obtain a limit at somewhat lower energies which are currently better investigated (e.g. $\sim1$ TeV).\\
Let`s suppose accelerated protons interact with low energy protons ($pp$ interaction) and produce \grays. If the source is transparent for produced \grays, one can derive  a limit on the \gray flux which accompanying HE \nray flux can be detected by current instruments. The sensitivity of nowadays HE \nray detectors (e.g. IceCube) at 1 TeV are:
\begin{equation}
F_{sens}(E_\nu=1\:TeV) = \left\{
  \begin{array}{lr}
    1.7\times10^{-11} \: \: \: \mathrm{TeV^{-1}cm^{-2}s^{-1}} & -\:\mathrm{after\:1\:year}\\
    4.9\times10^{-12} \: \: \: \mathrm{TeV^{-1}cm^{-2}s^{-1}} & -\: \mathrm{after\:5\:years}
  \end{array}
\right.
\label{eq1}
\end{equation}
On the other hand, the ratio between \gray and \nray fluxes produced in $pp$ interaction for broad proton distribution ($dN_{p}/dE_p\sim E_{p}^{-\alpha}$, $\alpha=(1.5\div3.5)$) corresponds to: $\frac{F_\gamma}{F_{\nu}}\approx(0.6-2.5)$ (see Fig. 2 in \cite{kel}). Taking into account this ratio and assuming that the HE \nray flux corresponds to the threshold value presented in Eq. \ref{eq1}, it can be calculated the corresponding \gray luminosity for a given distance ($d$). This luminosity corresponds to:
 \begin{equation}
L_{\gamma}(E_\gamma=1\:TeV) = \left\{
  \begin{array}{lr}
    (1.95\div8.13)\times10^{33}\left(\frac{d}{1\:kpc}\right)^2 \: \: \: \mathrm{erg\:s^{-1}} & -\:\mathrm{after\:1\:year}\\
    (5.6\div23.3)\times10^{32}\left(\frac{d}{1\:kpc}\right)^2\: \: \: \mathrm{erg\:s^{-1}} & -\: \mathrm{after\:5\:years}
  \end{array}
\right.
\label{eq2}
\end{equation}
A \gray source can be also detected by current \nray detectors, if at the distance $d=1$ kpc it has a \gray luminosity exceeding the limits presented in the Eq. \ref{eq2}. Accordingly, measuring the \gray flux at 1 TeV (and a corresponding luminosity) by comparing with the values presented in Eq. \ref{eq2} we can find relevant sources for observations with HE \nray detectors. The limit presented in Eq. \ref{eq2} was obtained assuming the source is transparent for produced \grays. The absorption of the \grays introduces uncertainties, between \gray and \nray fluxes and the limits presented in Eq. \ref{eq2} are not applied anymore. In particular, such conditions are satisfied for galactic binary systems which we discuss separately. 
\section{The sample}
\label{sec-2}
In principle above mentioned limits can be applied for both galactic and extragalactic sources. Poor understanding of physics of  extragalactic sources (e.g. AGN and GRBs) and intrinsic and/or intergalactic absorption hardens the prediction for extragalactic objects. Instead, the situation is more clear with the galactic sources: comparatively better information on physical processes responsible for \gray and \nray emissions. The recent observations in GeV/TeV band clearly demonstrated that several classes of galactic sources, such as supernova remnants, pulsar wind nebulae and microquasars (binary systems), are effective particle accelerators above TeV energies. However, in some cases both leptonic and hadronic scenarios can successfully reproduce the observed HE \gray data. Without going into details, here we assume that the observed \grays have a hadronic origin and we discuss the perspectives of detecting HE neutrinos from these sources.\\
All the results from observations in HE energy band (including all instruments) are listed in \textit{TeVCat}. Currently it contains around 150 sources which include both galactic and extragalactic objects. We consider only sources belonging to SNRs and PWNe classes, which are powerful nonthermal emitters up to very high energies.\\
\textbf{Supernova Remnants:}
SNRs were considered as the main sources of galactic cosmic rays after the arguments by Baade and Zwicky in 1934 \cite{baade}. A modest efficiency of $\sim$ 10 \% in converting the kinetic energy of supernova shocks into particle acceleration and the rate of SNRs in our Galaxy can explain the observed flux of cosmic rays. There was no observation proof of proton acceleration in SNRs shock until recently, since both leptonic and hadronic modelings give a satisfactory description of the \gray data. However, recently AGILE and Fermi-LAT measured pion bumps in SNRs IC 443 and W 44  \cite{tav1,tav2,acker} which show indirect evidence of the acceleration of hadronic particles in SNRs. If so, it is natural to consider SNRs as potential targets for HE \nray observations.\\
\begin{figure}
\centering
\includegraphics[width=14cm,clip]{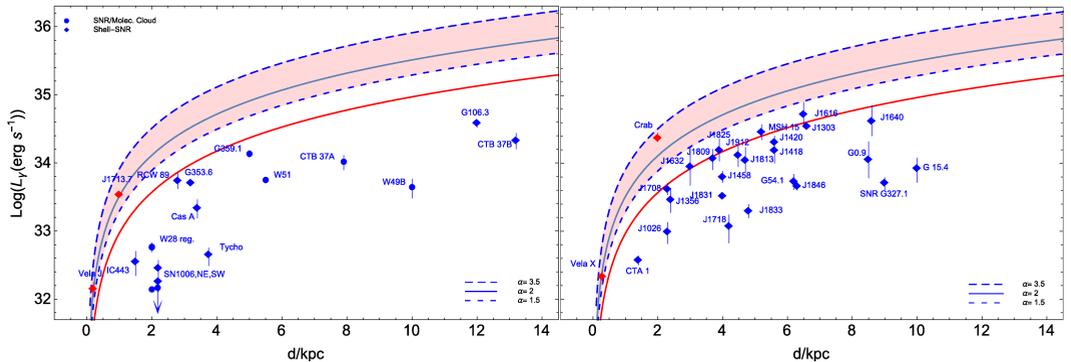}
\caption{Gamma-ray luminosity of SNRs (on the left) and PWNe (on the right) compared with the minimal \gray luminosity required to produce detectable HE \nray flux. The filled area corresponds to different power-law indicis of initial protons ($\alpha=(1.5\div3.5)$).}
\label{fig-1}       
\end{figure}
Currently, there are 10 SNRs interacting with molecular clouds and 13 shell type SNRs which are listed in \textit{TeVCat}. Unfortunately, for some sources there are missing measurements  of flux at 1 TeV (e.g. photon index) and only 20 SNRs are included in the sample. We assume, that the \grays have pure hadronic origin and we compare the luminosity at 1 TeV with limits obtained in Eq. \ref{eq2}. Fig. \ref{fig-1} (on the left) illustrates the luminosity of shell type SNRs (circle) and SNRs interacting with molecular clouds (square) compared with the \gray luminosity. The filled area corresponds to different initial proton distributions changing from very soft $\alpha=3.5$-long dashed to very hard $\alpha=1.5$-dashed (solid line represents the power-law index $\alpha=2$). As one can see, most of sources have \gray flux nearly two order of magnitude below than required threshold. It appears only for two powerful SNRs, the measured \gray luminosity is in the order of presented limit. These are SNRs  RX J1713.7-3946 and RX J0852.0-4622 which have been extensively discussed as potential targets for HE \nray detectors. Different estimations show that in principle several events in a year exposure can be detected if the produced \grays have hadronic origin (however, up to now the leptonic origin can not be disfavored). The location of these two SNRs is not covered by IceCube, but they are perfect targets for future planned KM3Net instrument.\\
The red solid line in Fig. \ref{fig-1} corresponds to necessary \gray luminosity in order to detect HE \nray -s after 5-years of observations. It is noticeable, that even in this case the situation is not improved: only above mentioned SNRs will be detected. The \gray observations should find new SNRs otherwise the expectations from currently known SNRs are not promising.\\
\textbf{Pulsar wind nebulae:}
PWNe are another class of galactic sources which accelerate particles up to VHEs. If one assumes the fraction of accelerated particles are protons, then from the inelastic interactions between accelerated protons and the ambient gas HE/VHE \grays and \nray -s are produced. Of course, the presence of bayronic component in pulsar winds (coexisting) with the leptonic one is still an open question and moreover the pulsar must efficiently convert rotational energy into proton acceleration. However, considering the difficulties with the leptonic modeling, the hadronic scenario remains attractive for \gray emission from PWNe. This makes PWNe another class of galactic sources producing HE \nray -s.\\  
The number of PWNe detected in TeV band and presented in \textit{TeVCat} corresponds to 35 but for 7 sources no data are available. For remaining 28 sources, we assume that the produced \grays have pure hadronic origin and compare with the limits presented in Eq. \ref{eq2} (see Fig. \ref{fig-1}). Similar to the previous case, the luminosity of almost all sources is significantly below than the required limit and only two sources, Crab nebulae and Vela X, have flux satisfying mentioned limit. One of these sources, crab nebulae - a very well studied object in all wavelengths, emits \grays up to TeV band which are most likely produced from inverse Compton scattering of soft photon fields inside the nebula. In case if \grays are produced from protons, from this source also HE \nray can be detected. However, so far no evidence for proton contribution seen from the observations.  On the other hand, Vela X PWN, an active pulsar (PSR B0833-45) associated with nebula, can be easily detectable with the current instrument (\gray flux corresponds to $L_\gamma\sim1.7\times10^{32}\:\mathrm{erg\:s^{-1}}$ exceeding the obtained limits). The origin of unusual TeV \grays (photon index $\Gamma\approx1.32$) \cite{abram} is under debates but considering the difficulties for modeling within leptonic scenarios, the hadronic origin can not be disfavored. All this arguments show that Vela region (Vela X PWN and Vela Junior SNR) is a prime candidate for galactic \nray sources to be detected with KM3Net.\\
It should be noted, that after 5-years of exposure, some of the sources can be detected with current HE \nray instruments (red solid line in Fig. \ref{fig-1}). Compared with the previous case, this is improved, but still few events in 5-years of observations do not make these source potentially interesting for \nray astronomy.\\
\textit{Binary systems:}
For a long time, binary systems were considered as potential \gray emitters, however unavoidable absorption of HE \grays (by optical photons in the systems) makes hard their detection. The improved sensitivity of current instruments results the detection  of GeV/TeV \grays from binary systems (microquasars) confirmed that their emission extends up to TeV energies (or even higher energies). If the \grays are produced from protons, the flux of HE \nray -s can significantly exceed the one of \grays (by a factor of $e^{\tau}$ where $\tau$ is the optical depth) considering that they escape from the region without absorption unlike the \grays. Due to the unknown parameter  $\tau$ no acceptable limit can be drown in this case. However, below it is demonstrated that using the observed \gray flux from Cygnus X-3 (well known galactic microquasar) and under reasonable assumptions for absorption it can be detected by neutrino detectors.\\
It has been shown, that \gray flaring activity from microquasar Cygnus X-3 can be well explained within hadronic scenario \cite{sah}: protons accelerated via jet interact with the wind particles from companion Wolf–Rayet star. Normalizing the proton spectrum using AGILE data (unabsorbed) and taking into account the absorption above GeV energies (in order not to overcome MAGIC upper limits) the expected HE \nray flux is only by a factor of about 3 lower than the 1-year IceCube sensitivity at $\sim$10 TeV.  Considering a prolonged "soft X-ray state" (which is characterized by \gray emission), the \nray flux from this source might be close to being detectable by cubic-kilometer \nray telescopes such as IceCube. Moreover, the estimations for HE \nray -s from another galactic binary system, LS 5039, show that the flux of HE \nray -s can be as large as $1.6\times10^{-11}\:\mathrm{cm^{-2}\:s^{-1}}$ (for energy greater than 1 TeV) \cite{ahar2}, above the sensitivity threshold of experiments in the Mediterranean Sea (KM3Net). This shows that probably binary systems are a class of galactic objects, from which the first galactic HE \nray -s will be detected.\\
\section{Conclusion}
The cubic kilometer scale HE \nray detectors with the existing HE \gray detectors provide an opportunity to study nonthermal processes in both galactic and extragalactic objects in great detail. While the recent detection of PeV \nray by IceCube demonstrate the beginning of extragalactic \nray astronomy, the first galactic sources of HE \nray still have to be detected. There are several classes of galactic sources, which \textit{i)} can accelerate the particles (hadrons) up to hundreds of TeV energies and \textit{ii)} are established to be TeV emitters. Clearly, the first galactic source which will be detected, belongs to one of these well established classes of TeV emitters.\\
The flux of TeV \gray allows to  estimate the expected flux of HE \nray, if the \grays have a hadronic origin and produced \grays do not suffer from absorption in the sources.  Indeed, the minimal detectable flux of HE \nray implies the minimal flux of accompanying  \grays. Such estimation shows, that from already detected SNRs and PWNe, only two SNRs, RX J1713.7-3946 and RX J0852.0-4622, one PWN- Vela X, have significantly high HE \nray flux, which can be detected with current \nray detectors. Even longer exposure time (5-years) will not significantly increase the number of potentially detectable sources. However, if the \grays from above mentioned sources have a hadronic origin they will provide several events  in kilometer-cube scale size detector in a year exposure.\\
On the other hand, the unexpected high flux of HE \nray -s can be detected from galactic binary systems (recently established strong TeV emitters). The effective absorption of TeV \grays, means that the flux of HE \nray can exceed the \gray one by several orders of magnitude (depending on the optical depth $\tau$). The estimation of HE \nray flux from several binary systems (e.g., Cygnus X-3,  LS 5039) shows that it is either in the same order or exceeds the sensitivity of the current instruments. Moreover, considering that the TeV \grays can be totally absorbed in the system (high density of soft photons), the \nray observations might discover new binary systems which are not "seen" by \gray instruments. Therefore, these are the best targets for HE \nray detectors.\\

\end{document}